\documentstyle[12pt]{article}
\def\fun#1#2{\lower3.6pt\vbox{\baselineskip0pt\lineskip.9pt
  \ialign{$\mathsurround=0pt#1\hfil##\hfil$\crcr#2\crcr\sim\crcr}}}
\relax
\advance\textheight by \topskip
\begin{document}

\newcommand{\DD}[2]{\mbox{$\frac{\partial^{#1}}{\partial #2}$}}
\newcommand{\Ddd}[2]{\mbox{$\frac{\partial^{2}}{\partial #1
\partial #2}$}}
\newcommand{\rx}{\mbox{$\rho(X)$}}
\newcommand{\kx}{\mbox{$K(X,Y)$}}
\newcommand{\uv}[2]{\mbox{$u(#1,#2,z)$}}
\newcommand{\Dme}{\mbox{$m_{1}(t,X)$}}
\newcommand{\EV}[1]{\mbox{${\bf{\sf E}}\left\{ \left. #1 \right| \,
X_{0}=X \right\} $}}
\newcommand{\PE}[1]{\mbox{${\bf{\sf P}} \left\{ \left. #1 \right| \,
X_{0}=X \right\} $}}
\newcommand{\Pe}[2]{\mbox{${\bf{\sf P}}_{#1}\left\{ #2 \right\} $}}
\newcommand{\Pb}[2]{\mbox{$\overline{{\bf{\sf P}}}_{#1}\left\{ #2
\right\} $}}
\newcommand{\Pw}[3]{\mbox{${\bf{\sf P}}\left\{ \left. #1 \right| \,
X_{#2}=X, \, X_{#3}=Y \right\} $}}
\newcommand{\Pes}{\mbox{$\psi_{1}(X) $}}
\newcommand{\Dpi}{\mbox{$\pi_{1}(X) $}}
\newcommand{\SD}[2]{\mbox{${\rm SD}_{#1}^{#2}$}}
\newcommand{\SC}[2]{\mbox{${\rm SC}_{#1}^{#2}$}}
\newcommand{\Qd}[3]{\mbox{$Q_{#1,#2}(#3)$}}
\newcommand{\man}{\mbox{${\cal M}\backslash \Gamma$}}
\newcommand{\IK}[1]{\mbox{$\int_{{\cal M}\backslash \Gamma} \, dY
\kx \, \Qd{D}{k}{#1}   $}}
\newcommand{\IKB}[1]{\mbox{$\int_{{\cal M}\backslash \Gamma}
\kx \: \Qb{D}{k}{#1} \, dY  $}}
\newcommand{\Qb}[3]{\mbox{$\overline{Q}_{#1,#2}(#3) $}}
\newcommand{\isu}[4]{\mbox{$\int_{#1}^{#2} du
\frac{\sinh(#3)}{\sinh(\zeta)} \, #4 $}}
\newcommand{\rot}{\mbox{$\rho (X,t)$}}
\newcommand{\Kx}{\mbox{$K(X,Y,t)$}}
\newcommand{\KX}[2]{\mbox{$K(#1,Y,#2)$}}
\newtheorem{prop}{Proposition}
\newtheorem{theorem}{Theorem}
\newtheorem{defi}{Definition}

\rightline{SU-ITP-92-32}

\rightline{October 28, 1992}

\vspace{1cm}

\begin{center}
\Large{INFINITE-SCALE PERCOLATION IN A NEW TYPE \\
OF BRANCHING DIFFUSION PROCESSES}   \\
\vspace{1cm}
{\bf A.Mezhlumian}\mbox{$\phantom{.}^{\dag}$}\footnote{
On leave from L.D.Landau Institute for Theoretical Physics,
Moscow, Russia.}, \hspace{0.5cm}
{\bf S.A.Molchanov}\mbox{$\phantom{.}^{\ddag}$} \\
\vspace{1.5cm}
\mbox{$\phantom{.}^{\dag}$}
{\it Department of Physics, Stanford University, \\
Stanford, CA 94305-4060 \phantom{.}\footnote{
E-mail: arthur@physics.stanford.edu}} \\
\vspace{1cm}
\mbox{$\phantom{.}^{\ddag}$}
{\it Department of Mathematics, University of California \\
Irvine, CA 92717 , USA }
\end{center}

\vspace{2cm}
{\centerline {\large  ABSTRACT}}

\vspace{1cm}

We give an account of matter and (basically) a solution of a new
class of problems synthesizing percolation theory and branching
diffusion processes. They led us to realizing a novel type of
stochastic processes, namely  branching processes with  diffusion
on the space of parameters distinguishing the branching ``particles''
each other.

\newpage
\section{INTRODUCTION}        \label{sec1}

Branching stochastic processes \cite{Sev71} have always been very
interesting for mathematicians and physicists. They describe well
a multitude of phenomena from chain reactions to populational
dynamics. On the other hand, ordinary \cite{Kest82} and multi-scale
\cite{MMS86} percolation play a crucial role in many applications
\cite{ShK71}. In the work \cite{MPR88} (see also \cite{MPR87})
on the basis of the idea of recoding \cite{ALMP87} necessary and
sufficient conditions for the discrete hierarchical (multi-scale)
model of fracture (percolation of defects) were obtained.

In this paper we give an account of matter and (basically) a solution
of a new class of problems synthesizing percolation theory
and branching diffusion processes. Such problems arose naturally
in the recent investigations of the global
geometry of inflationary early Universe (see \cite{LinM91}
and references therein).

In Section \ref{sec2} we present a general formulation of  problem
of infinite-scale percolation in a new type of branching diffusion
processes. The novelty is the diffusion not in the real space where
branching ``particles'' live, but in the space of parameters
distinguishing the ``particles''  each other.

We do not specify in the following, whether the $A$-diffusion on a
$d$-dimensional manifold, which we assume from the very beginning,
corresponds to some stochastic differential equation (SDE) in the
sense of It\^{o} or in the sense of Stratonovich. In the
applications~\cite{LinM91}
we use the Stratonovich's symmetrized calculus
because white noise there is the limit of a colored one with small
time correlation. The generating operator
($\hat{A}$-operator) of the diffusion in local coordinate frame
looks like \cite{IW81} ($i,j,k=1,2,\ldots ,d $ ; the summation over
repeated indexes is assumed)

{\bf i)}    after It\^{o}

\begin{equation}
 \hat{A_{I}} f(X)=\frac{1}{2} \sigma^{ik}(X) \sigma^{kj}(X)
 \Ddd{X^{i}}{X^{j}} f(X) + {\cal F}^{i}(X) \DD{}{X^{i}} f(X)
                                       \label{eq1.1}
\end{equation}

{\bf ii)}     after Stratonovich

\begin{equation}
 \hat{A_{S}} f(X)=\frac{1}{2} \sigma^{ik}(X) \DD{}{X^{i}}
 \left( \sigma^{kj}(X) \DD{}{X^{j}} f(X) \right)  +
 {\cal F}^{i}(X)  \DD{}{X^{i}}   f(X)
                                       \label{eq1.2}
\end{equation}

\noindent and corresponds to SDE (in the proper sense)

\begin{equation}
 dX^{i}_{t} = {\cal F}^{i}(X)\, dt + \sigma^{ik}(X) \circ dW^{k}_{t}
                                                 \label{eq1.3}
\end{equation}

Here $X^{i}_t \in R^{d}, W^{k}_{t}$ is the $d$-dimensional Wiener
process. We add to the random walk~(\ref{eq1.3}) a branching
with intensity $n(X)$ (i.e.\  the probability density of branching
of the ``particle'' at $X$ during the time $\Delta t$ is
$n(X)   \Delta t$ ). Then we relate with such a branching diffusion
some process of breaking up of $D$-dimensional cubes and
their coloring and obtain an infinite-scale (in the
$t \rightarrow \infty $ limit) percolation problem.

After some preliminary (Section \ref{sec3}) the solution of the
problem is stated in Section \ref{sec4}  where necessary and
sufficient conditions for percolation are obtained.
In Section \ref{sec5} we illustrate our method in the case
of the simplest model. In the Summary (Section \ref{sec6}) possible
generalizations are discussed.


\section{THE PROBLEM}                     \label{sec2}

Let the basic $d$-dimensional diffusion process
generated by the operator
$\hat{A}$ (\ref{eq1.1}), (\ref{eq1.2})  given

\begin{equation}
\lim_{\Delta t \downarrow 0} {\bf{\sf E}}
\left\{ \left. \frac{f(X_{t+\Delta t}) - f(X_{t})}{\Delta t} \right|
\,
X_{t}=X \right\}
= \hat{A} f(X)
                                             \label{eq2.1}
\end{equation}

\noindent where $X_{t} \in {\cal M}$, ${\cal M}$
is a $d$-dimensional manifold.
An action of $\hat{A}$ is endowed by the boundary condition

\begin{equation}
 f(X)|_{X \in \Gamma} = 0                       \label{eq2.2}
\end{equation}

\noindent where $\Gamma \subset {\cal M} $
is a closed subset of ${\cal M}$
(an absorbing boundary).

On ${\cal M} \backslash \Gamma $  smooth functions
$n(X), {\cal F}^{i}(X), \sigma^{ij}(X) $ are defined
(see (\ref{eq1.3})), where $n(X)$ and $\sigma^{ij}(X) $
are assumed to be positive,
such that $\Gamma$ is accessible.
Let in each branching instead of one ``particle'' at point $X$
appear the given number $r$ of ``particles'' at that point, which
continue to evolve as branching diffusion
process independently from each other.
When a ``particle'' reaches the absorbing boundary $\Gamma$,
it stays there forever without branching.

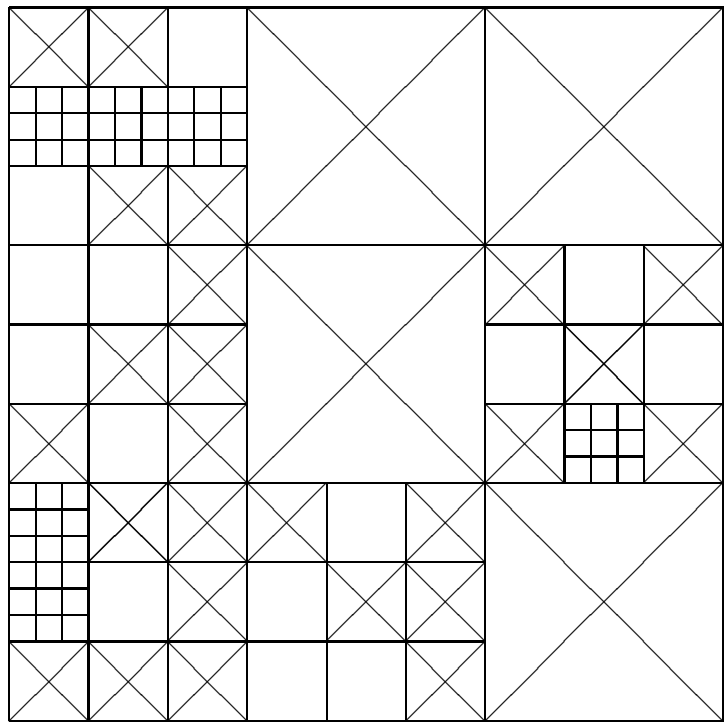
\begin{figure}[p]
\phantom{.} \vfill
\phantom{.} \hfill
\begin{picture}(270,270)
\multiput(0,0)(90,0){4}{\line(0,1){270}}
\multiput(0,0)(0,90){4}{\line(1,0){270}}
\multiput(0,0)(0,30){4}{\line(1,0){180}}
\multiput(0,0)(30,0){4}{\line(0,1){270}}
\multiput(0,90)(0,30){7}{\line(1,0){90}}
\multiput(120,0)(30,0){3}{\line(0,1){90}}
\multiput(180,90)(0,30){4}{\line(1,0){90}}
\multiput(180,90)(30,0){3}{\line(0,1){90}}
\multiput(180,0)(0,90){3}{\line(-1,1){90}}
\multiput(90,90)(0,90){2}{\line(1,1){90}}
\multiput(180,0)(0,180){2}{\line(1,1){90}}
\multiput(270,0)(0,180){2}{\line(-1,1){90}}
\multiput(150,0)(-30,30){3}{\line(1,1){30}}
\multiput(150,30)(0,30){2}{\line(1,1){30}}
\multiput(180,30)(0,30){2}{\line(-1,1){30}}
\multiput(0,0)(30,60){3}{\line(1,1){30}}
\multiput(30,0)(30,60){3}{\line(-1,1){30}}
\multiput(60,0)(0,30){4}{\line(1,1){30}}
\multiput(90,0)(0,30){4}{\line(-1,1){30}}
\multiput(30,0)(0,60){5}{\line(1,1){30}}
\multiput(60,0)(0,60){5}{\line(-1,1){30}}
\multiput(0,90)(60,60){2}{\line(1,1){30}}
\multiput(30,90)(60,60){2}{\line(-1,1){30}}
\multiput(60,180)(-60,60){2}{\line(1,1){30}}
\multiput(90,180)(-60,60){2}{\line(-1,1){30}}
\multiput(180,90)(30,30){3}{\line(1,1){30}}
\multiput(210,90)(30,30){3}{\line(-1,1){30}}
\multiput(240,90)(-30,30){3}{\line(1,1){30}}
\multiput(270,90)(-30,30){3}{\line(-1,1){30}}
\multiput(10,30)(10,0){2}{\line(0,1){60}}
\multiput(0,40)(0,10){5}{\line(1,0){30}}
\multiput(10,210)(10,0){8}{\line(0,1){30}}
\multiput(0,220)(0,10){2}{\line(1,0){90}}
\multiput(220,90)(10,0){2}{\line(0,1){30}}
\multiput(210,100)(0,10){2}{\line(1,0){30}}
\end{picture} \hfill \phantom{.}
\caption[A part of the net, containing black squares of
different sizes in a sea of white squares. Crossed
squares are black.]{ \label{fig1}}
\phantom{.} \vfill
\end{figure}

Now we will associate this branching diffusion with the following
picture. Let at $t=0$ a $D$-dimensional cubic net be given,
consisting
of cubes of unit size. With each cube of this net independently we
set in correspondence a random point $X_{0} \in {\cal M}$ with the
probability density $\gamma (X_{0})$. The subsequent evolution of
each
cube is independent from the rest of the net. This evolution is
determined by the abovedescribed
basic branching diffusion on ${\cal M}$\@.
At $t=0 \; \: r$ trajectories $X^{(i)}_{t}
\; (i=1,\ldots,r)$ of random walkers
(\ref{eq1.3}) start at $X_{0}$. We divide the unit cube (call it
cube of zero level) into $r$ smaller equal cubes of first
level (we assume that $r=k^{D}$, so that the cubes of first level are
$k$ times smaller than the unit cube). Each first level cube is set
in
correspondence with one of the points~$X^{(i)}_{t}$ --- the points
where the $i$-th ``particle'' branches
for the first time. If the ``particle''
was absorbed at $\Gamma$, we attribute to the corresponding first
level cube the proper point $X_{1} \in \Gamma$, color it black
and leave it in peace. We divide the other cubes, which
should not be colored this
time (we call them ``white'' or ``living'' cubes), once more
into the $r=k^{D}$ cubes of second level and do the same procedure,
starting from $X^{(i)}_{1} \in {\cal M} \backslash \Gamma $.
Thus, coloring some cubes at each level $q$ (when
$X^{(i)}_{q} \in \Gamma$ ) and dividing white cubes further
we will obtain some infinite-scale
(provided that the process does not
degenerate at finite level) picture of black cubes of different
sizes in a ``sea'' of living white cubes (see Fig. \ref{fig1}).

Denote ${\cal B}_{t}$ the multitude of all black cubes of all
sizes (and ${\cal W}_{t}$ --- of all white cubes) belonging
to a particular unit cube at the time $t$.
We assume that two cubes of (possibly) different sizes are
connected if they have common face. Now we are ready to
formulate the

\begin{description}
\item[Problem:] Consider the whole net of (zero level) cubes
and denote by ${\sf B}_{t}$ the union of all the multitudes
${\cal B}_{t}$ belonging to all unit cubes in the original
cubic net. Let $\gamma(X)$, $n(X)$
and $\hat{A}$ be given. Does ${\sf B}_{t}$ percolate
in the $t \rightarrow \infty$ limit (we will denote the
corresponding multitudes of cubes as ${\cal B}_{\infty}$
and ${\sf B}_{\infty}$ respectively)?
\item[Note:] Percolation of \mbox{${\sf B}_{\infty}$} means that
a connected non-selfintersecting path  a.\ s.\
exists on \mbox{${\sf B}_{\infty}$} from some
point (call it origin of the coordinate frame) up to infinity.
\end{description}

\section{PRELIMINARY}             \label{sec3}

First of all, let us elucidate the condition of non-degeneracy
of the process of breaking up of the cubes.
Let $\mu(t,V)$ be the number of random walkers within the
region $V \subset {\cal M} \backslash \Gamma $ at the time
$t$. Introduce the generating function

\begin{equation}
\uv{t}{X} = \EV{z^{\mu(t,V)}}         \label{eq3.1}
\end{equation}

\noindent Noting that the evolution of \uv{t}{X} is driven by the
diffusion
of particles and by the branching processes, we obtain

\begin{equation}  \label{eq3.1.1}
\uv{t+\Delta t}{X} = (1 - n(X) \, \Delta t) \, \EV{\uv{t}{X_{\Delta
t}}}
+ n(X) \, \Delta t \, (\uv{t}{X})^{r}
\end{equation}

Here we considered the whole tree of the branching diffusion
process from $t=0$ to $t + \Delta t$ and divided it into two parts
---
a part from $t=0$ to $t = \Delta t$ and a part from $t = \Delta t$ to
$t + \Delta t$\@. Up to the first order in $\Delta t$, we have only
two mutually exclusive possibilities of evolution of the process
from $t=0$ to $t = \Delta t$\@.
\begin{enumerate}
\item The single original particle does not branch during this time
and
it diffuses to new place $X_{\Delta t}$\@. The tree of the branching
diffusion process from $t = \Delta t$ to $t + \Delta t$ differs from
the
one corresponding to the interval of time $[0,t]$ only by its
origin $X_{\Delta t}$\@. This possibility is represented by the first
term in the r.h.s.\  of Eq.(\ref{eq3.1.1})\@.
\item The original particle branches just one time during the period
from $t=0$ to $t = \Delta t$\@. There appear $r$ species of the tree
of the branching diffusion process and,
in the first order in $\Delta t$, we should not distinguish them
from the tree corresponding to the interval $[0,t]$\@.
This possibility is represented by the
second term in the r.h.s.\  of Eq.(\ref{eq3.1.1})\@.
\end{enumerate}

\noindent Then, using Eq.(\ref{eq2.1}),
one easily derives from (\ref{eq3.1.1})
the (backward) differential equation for the generating function

\begin{equation}
\DD{}{t} \uv{t}{X} = \hat{A} \, \uv{t}{X} +
n(X) \, (u^{r}(t,X,z) - \uv{t}{X})        \label{eq3.2}
\end{equation}

\noindent and the boundary and initial conditions

\begin{equation}
\uv{t}{X} |_{X \in \Gamma} =1       \label{eq3.3}
\end{equation}

\begin{equation}
\uv{0}{X} = \left\{ \begin{array}{ll}
                      z, & \mbox{if $X \in V$} \\
                      1, & \mbox{otherwise}
                      \end {array}  \right.
                                    \label{eq3.4}
\end{equation}

As a consequence of the definition (\ref{eq3.1}) we can write
down the equations for the (factorial) moments of $\mu(t,V)$
differentiating the Eqs.\  (\ref{eq3.2})-(\ref{eq3.4}).
Note that

\begin{equation}
m!_{l}(t,X) = \EV{\mu (\mu -1) \ldots (\mu -l)}
= \DD{l}{z^{l}} \uv{t}{X}|_{z=1}
                                       \label{eq3.5}
\end{equation}

\noindent and

\begin{equation}
m_{l}(t,X) = \EV{\mu^{l}(t,V)} = \left( z \DD{}{z}
\right)^{l} \uv{t}{X}|_{z=1}
                  \label{eq3.6}
\end{equation}

\noindent In particular, the equation for the first moment
$\Dme = z \DD{}{z} \uv{t}{X}|_{z=1} $ (the average number of
random walkers in $V$ at the time $t$ ) is

\begin{equation}
\DD{}{t} \Dme = \hat{A} \, \Dme +
(r-1) \, n(X) \, \Dme      \label{eq3.8}
\end{equation}

\begin{equation}
\Dme |_{X \in \Gamma} = 0  \label{eq3.9}
\end{equation}

\begin{equation}
m_{1}(0,X) = \left\{ \begin{array}{ll}
                 1, & \mbox{if $ X \in V$} \\
                 0, & \mbox{otherwise}
                 \end{array}    \right.
                        \label{eq3.10}
\end{equation}

\noindent Eqs.\  (\ref{eq3.8})-(\ref{eq3.10}) have an asymptotic
($t \rightarrow \infty $) solution

\begin{equation}
\Dme \propto e^{\lambda_{1} t} \psi_{1}(X)
\int_{V} dY   \pi_{1}(Y)
        \label{eq3.11}
\end{equation}

\noindent Here \Pes \, is the unique strictly positive
real eigenfunction of the marginal problem

\begin{equation}
\hat{A} \, \Pes + (r-1) \, n(X) \, \Pes = \lambda_{1} \Pes
\label{eq3.12}
\end{equation}

\begin{equation}
\Pes |_{X \in \Gamma} = 0 \label{eq3.13}
\end{equation}

\noindent and $ \lambda_{1} $ \, is the
corresponding (real) eigenvalue.
The function \Dpi \, \, (invariant density) is the unique strictly
positive eigenfunction of the adjoint equation with the same
eigenvalue~$\lambda_{1} $

\begin{equation}
\mbox{$\hat{A}^{\dag}$} \, \Dpi + (r-1) \, n(X) \, \Dpi =
\lambda_{1} \Dpi
                  \label{eq3.14}
\end{equation}

\begin{equation}
\Dpi |_{X \in \Gamma} =0  \label{eq3.15}
\end{equation}

\noindent The normalizations are as follows

\begin{equation}
\int_{{\cal M} \backslash \Gamma} \Dpi \, dX  = 1 \; \; ; \; \;
\int_{{\cal M} \backslash \Gamma} \Dpi \, \Pes \, dX = 1
\label{eq3.16}
\end{equation}

One can see from (\ref{eq3.11}) that if $\lambda_{1} < 0 $, then
the branching process a.\ s.\  degenerates in the
limit $t \rightarrow \infty $
(in our notations ${\cal W}_{\infty} = \emptyset $).
If $\lambda_{1} > 0 $,
then the branching process is supercritical and
${\cal W}_{\infty} \neq \emptyset $ . We assume the latter case in
the rest of the paper.

Let us introduce two other useful functions. The first
one is the probability of the event $\aleph_t$ that
the particle gets absorbed at $\Gamma$
at some time less or equal $t$ provided that it started to diffuse
from some point $X \in {\cal M} \backslash \Gamma $
at the time $t = 0$ and didn't branch before being absorbed.

\begin{equation}
\rot = \PE{\aleph_t}
\end{equation}

\noindent Consider the moment of time $t + \Delta t$\@.
It is obvious from the
definition above that (in the first order in $\Delta t$)

\begin{equation}   \label{21.2}
\rho(X,t + \Delta t) = (1 - n(X) \, \Delta t) \,
\EV{\rho(X_{\Delta t},t)}
\end{equation}

\noindent and the following (backward) differential
equation as well as the initial
and boundary conditions follow immediately from (\ref{21.2})
and (\ref{eq2.1})

\begin{equation}
\DD{}{t} \rot = \hat{A} \, \rot - n(X) \, \rot      \label{eq3.17}
\end{equation}

\begin{equation}
\rot |_{X \in \Gamma} = 1         \label{eq3.18}
\end{equation}

\begin{equation}
\rho(X,0) = \left\{ \begin{array}{ll}
                 1, & \mbox{if $ X \in \Gamma$} \\
                 0, & \mbox{otherwise}
                 \end{array}    \right.
                        \label{23.1}
\end{equation}

In the rest of this paper we will use only the stationary
probability $\rho(X)$ of absorption of a particle, which
starts its random walk at $X$, without branching before absorption.
It is given by the stationary solution
of Eqs.\ (\ref{eq3.17})-(\ref{23.1})\@.
Since $n(X)$ is positive and the maximal eigenvalue of $\hat{A}$ with
zero boundary condition on $\Gamma$ is negative, $\lambda = 0$ is
not an eigenvalue of $(\hat{A} - n(X) )$ and the stationary equation

\begin{equation}
\hat{A} \, \rx - n(X) \, \rx = 0 \; \; ; \; \;
\rx |_{X \in \Gamma} = 1  \label{eq3.19}
\end{equation}

\noindent has unique solution. Of course, the relation
$\lim_{t \rightarrow \infty} \rot = \rx $  is satisfied.

Then, let us introduce the probability density in $Y$

\begin{equation}
\Kx = \PE{\Re_t(Y)} \label{eq3.19.1}
\end{equation}

\noindent where $\Re_t(Y)$ is the event that the
particle branches for the first
time in the infinitesimal volume $Y+dY$ at some time less or equal
$t$,
provided that it starts to diffuse at $X$ at $t = 0$\@.
Consider the evolution of this quantity after time $\Delta t$ in the
first order in $\Delta t$\@. One has

\begin{equation}   \label{28}
\KX{X}{t} = n(X) \, \Delta t \, \delta(Y-X) + (1 - n(X) \, \Delta t)
\,
\EV{\KX{X_{\Delta t}}{t + \Delta t}}
\end{equation}

Here the first term represents the event that the particle branches
during the time interval $[0, \Delta t]$ and we should not
distinguish
the position of the particle at $t=0$ and $t=\Delta t$ in the first
order in $\Delta t$\@. The second term represents the event that
the particle does not branch during the time interval $[0, \Delta t]$
and in the first order in $\Delta t$ the only difference we should
take
into account for that particle's trajectory during the time interval
$[\Delta t, t + \Delta t]$ is its random initial point
$X_{\Delta t}$ at the beginning of that interval. These
are the only two mutually exclusive events which exist
in the first order in
$\Delta t$\@. From the definition (\ref{eq3.19.1})
and Eq.\ (\ref{28}) one easily
obtains the (backward) differential equation and the boundary
and initial conditions

\begin{equation}
\DD{}{t} \Kx = \hat{A}_{X} \Kx - n(X) \, \Kx +n(Y) \, \delta (X-Y)
\label{eq3.20}
\end{equation}

\begin{equation}
\Kx |_{X \in \Gamma} = \Kx |_{Y \in \Gamma} = 0 \label{eq3.21}
\end{equation}

\begin{equation}
K(X,Y,0) = \delta (X-Y)  \label{eq3.22}
\end{equation}

Here $\hat{A}_{X}$ denotes the generating differential operator of
diffusion acting on the first variable of $K(X,Y,t)$\@. In what
follows
we will need only the stationary probability density $K(X,Y)$,
which satisfies the equations

\begin{equation}
\hat{A}_{X} \kx - n(X) \, \kx + n(Y) \, \delta (X-Y) = 0
\label{eq3.24}
\end{equation}

\begin{equation}
\kx |_{X \in \Gamma} = \kx |_{Y \in \Gamma} = 0    \label{eq3.25}
\end{equation}

One can write down the expression for \kx \, \, through the complete
orthonormal set of eigenfunctions $K_{s}(X)$ and eigenvalues
$\kappa_{s}$ of the marginal problem

\begin{equation}
\hat{A} \, K_{s}(X) - n(X) \, K_{s}(X) = \kappa_{s} K_{s}(X)
\label{eq3.26}
\end{equation}

\begin{equation}
K_{s}(X)|_{X \in \Gamma} = 0   \label{eq3.27}
\end{equation}

\noindent Recalling the following properties of the eigenfunctions

\begin{equation}
\int_{{\cal M}\backslash \Gamma} K_{i}(X) \, K_{j}(X) \, dX =
\delta_{ij} \; \; ; \; \;
\sum_{s=1}^{\infty} K_{s}(X) \, K_{s}(Y) = \delta(X-Y)
\label{eq3.28}
\end{equation}

\noindent one easily obtains

\begin{equation}
\kx = -n(Y) \sum_{s=1}^{\infty} \frac{1}{\kappa_{s}} K_{s}(X) \,
K_{s}(Y)
\label{eq3.29}
\end{equation}

Now, with the functions \rx \, \, and \kx \, \, in hand,
we are ready to solve the
problem stated in the preceding section.

\section{INFINITE-SCALE PERCOLATION}          \label{sec4}

In this section we will find the percolation characteristics
of \mbox{${\sf B}_{\infty}$} using the renormalization relations,
which arise as a result of the
recoding procedure. This procedure was introduced in the case of the
simpler model of discrete hierarchical fracture in \cite{ALMP87}
and was investigated in details in \cite{MPR88}.

Consider the multitude \mbox{${\cal B}_{t}$} , \, belonging
to a particular unit cube, at some large time~$t$\@.
Let us concentrate for the moment on the
percolation characteristics of this unit cube.
Let us define the notion of
``strongly defective'' cubes\footnote{We
call black cubes ``defects'' to
resemble the rock fracture theory \cite{ALMP87}\@.}\@.

\begin{defi}
A cube of level $L$ is called ``strongly defective'' of rank $0$
\mbox{\rm (\SD{L}{(0)})} if it is black.
It is called ``strongly defective''
of rank $m>0$ \mbox{\rm (\SD{L}{(m)})}, if
either it is black or the following
two conditions are satisfied:
\begin{enumerate}
\item There exists a connected cluster of \SD{L+1}{(m-1)} cubes
of level $L+1$ which connects each pair of faces of the given cube;
\item Strictly more than a half of the surface of
each face of the cube
is covered by \SD{L+1}{(m-1)} cubes belonging to one of the clusters
defined in 1)
\end{enumerate}
A cube is said to be \SD{}{} if it is \SD{}{(n)} for some
$n \geq 0$\@.
\label{def1}
\end{defi}

We assume in the main body of the text $k \geq 3$. In the $k=2$ case
more care is needed due to the fact that there are no nontrivial
configurations satisfying condition 2 (see Section \ref{sec6})\@.
The meaning of this definition is revealed by the following obvious

\begin{prop}
If any number of \SD{}{} cubes of some level $L$ are attached face
to face, then there is a connected cluster of black cubes
(consisting,
maybe, of cubes of different levels $K \geq L$) running
through them. If a cube
is \SD{}{(m)}, then it is \SD{}{(n)} for every $n>m$\@.
\label{prop1}
\end{prop}

Note, that it is the condition 2 of the Definition \ref{def1} that
ensures the percolation of the black color through the attached faces
of SD cubes. Thus, SD cubes are as good as black ones when we deal
with
the percolation of the black color through the net of cubes. It is
this kind of substitution of black cubes by
SD cubes that was called  ``recoding procedure'' in \cite{ALMP87}
and that enables us to follow the ``renormalization approach''
of \cite{MPR88,ALMP87}\@.

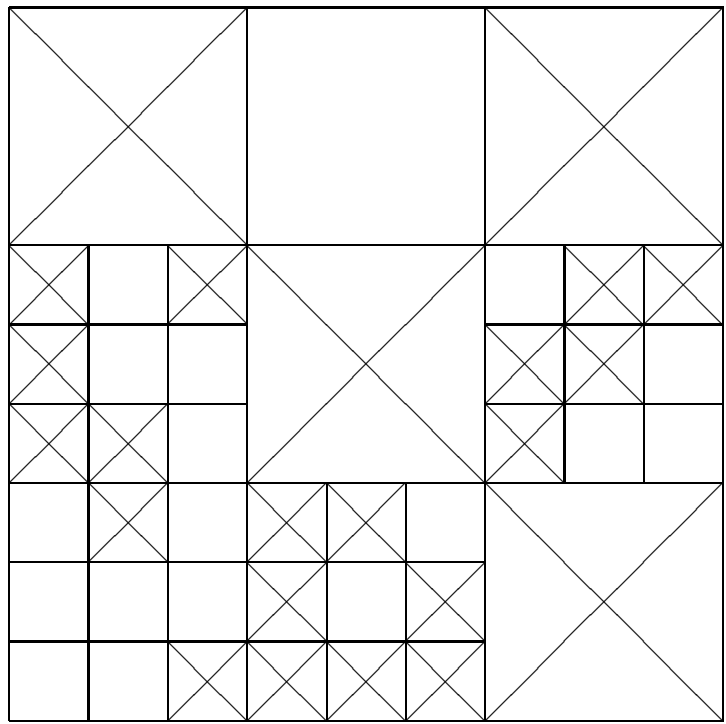
\begin{figure}[p]
\phantom{.} \vfill
\phantom{.} \hfill
\begin{picture}(270,270)
\multiput(0,0)(90,0){4}{\line(0,1){270}}
\multiput(0,0)(0,90){4}{\line(1,0){270}}
\multiput(0,0)(0,30){4}{\line(1,0){180}}
\multiput(0,0)(30,0){4}{\line(0,1){180}}
\multiput(0,90)(0,30){4}{\line(1,0){90}}
\multiput(120,0)(30,0){3}{\line(0,1){90}}
\multiput(180,90)(0,30){4}{\line(1,0){90}}
\multiput(180,90)(30,0){3}{\line(0,1){90}}
\multiput(0,180)(90,-90){3}{\line(1,1){90}}
\multiput(90,180)(90,-90){3}{\line(-1,1){90}}
\multiput(180,180)(0,-90){2}{\line(1,1){90}}
\put(270,180){\line(-1,1){90}}
\multiput(210,90)(30,30){3}{\line(-1,1){30}}
\multiput(180,120)(30,30){2}{\line(1,1){30}}
\multiput(210,120)(30,30){2}{\line(-1,1){30}}
\multiput(90,30)(0,30){2}{\line(1,1){30}}
\multiput(120,30)(0,30){2}{\line(-1,1){30}}
\multiput(60,0)(30,0){4}{\line(1,1){30}}
\multiput(90,0)(30,0){4}{\line(-1,1){30}}
\multiput(120,60)(30,-30){2}{\line(1,1){30}}
\multiput(150,60)(30,-30){2}{\line(-1,1){30}}
\multiput(0,90)(0,30){3}{\line(1,1){30}}
\multiput(30,90)(0,30){3}{\line(-1,1){30}}
\multiput(30,60)(0,30){2}{\line(1,1){30}}
\multiput(60,60)(0,30){2}{\line(-1,1){30}}
\put(60,150){\line(1,1){30}}
\put(90,150){\line(-1,1){30}}
\end{picture} \hfill \phantom{.}
\caption[Counterexample. The square of zero level is SD, although
four of nine squares of first level are not SD. ]{ \label{fig2}}
\phantom{.} \vfill
\end{figure}

An opposite proposition (that SD cubes always consist
of SD cubes) is false (see Fig.\ref{fig2}). That's why we can
obtain only sufficient (but not necessary) condition for percolation
assuming that the probability of level $L$ cube to be SD can be
computed neglecting such configurations as in Fig.\ref{fig2}\@

\begin{equation}
\Pe{X}{\SD{L}{}} = \Pe{X}{\mbox{\cal B}_{L}} +
(1-\Pe{X}{\mbox{\cal B}_{L}}) \:
\Pe{X}{\mbox{$\begin{array}{c}
     \mbox{SD-configuration of} \\
     \mbox{level \( L+1 \) cubes}
     \end{array}  $}}      \label{eq4.1}
\end{equation}

\noindent where the subscripts $L$ denote the
level of the cubes and the subscripts
$X$ take into account the fact that we deal with probabilities
depending
on the value of the parameter associated with the given cube. We are
interested  in calculation of \Pe{X}{\SD{0}{}}\@. Let us remind that
SD
means $\mbox{SD}^{(n)}$ for some $n \geq 0$\@. In other words, we
should
find the limit

\begin{equation} \label{4.1.1}
\Pe{}{X} = \lim_{n \rightarrow \infty} \Pe{X}{\SD{0}{(n)}}
\end{equation}

\noindent Let us forget for a while about the
time dependence of the problem
and use the stationary equation

\begin{equation}
\Pe{X}{\SD{0}{(n)}} = \rx + (1-\rx )\int_{{\cal M}\backslash \Gamma}
\kx \, \Qd{D}{k}{\Pe{Y}{\SD{1}{(n-1)}}} \, dY           \label{eq4.2}
\end{equation}

\noindent where \rx \, \, and \kx \, \, were
introduced in the preceding section
(see (\ref{eq3.19}),
(\ref{eq3.24})-(\ref{eq3.29}))\@. Equation (\ref{eq4.2}) gives the
probability for the zero level cube to be \SD{0}{(n)} in terms of
the probabilities of the first level cubes to be \SD{1}{(n-1)}
provided
that the parameters associated with them are such that the
``particle''
corresponding to the zero level cube is the ``parent'' of the
particles corresponding to the first level cubes.

The combinatorial function
\Qd{D}{k}{p} which appeared in (\ref{eq4.2}) is a polynomial
in $p$ of $r$-th degree. It counts the probability of SD
configurations of level $L+1$ cubes inside of cubes of
level $L$ (we assume for the moment that $p$ is the probability
of black color and $q=1-p$ is the probability of white one)\@.
We will consider in this paper only the simplest case of two
dimensional cubes (squares) and $k=3$, although it is only a
straightforward combinatorial problem to find \Qd{D}{k}{p} for the
other cases\footnote{One should remember, however, that computational
difficulties grow exponentially fast with growing $D$ or $k$, and
the finding of the functions \Qd{D}{k}{p}, although
remaining a finite combinatorial problem, becomes very
cumbersome.}\@.

\begin{equation} \label{36.1}
\Qd{2}{3}{p}=p^{9}+9p^{8}q+20p^{7}q^{2} = 12p^9 - 31p^8 +20p^7
\end{equation}

\noindent Equation (\ref{eq4.2}) can be continued by the hierarchy

\begin{eqnarray} \label{eq4.6}
\Pe{X}{\SD{1}{(n-1)}}=\rx + (1-\rx ) \,
\int_{\man} \, dY \kx \: \Qd{D}{k}{\Pe{Y}{\SD{2}{(n-2)}}}  \\
\vdots \nonumber \\
\Pe{X}{\SD{n-1}{(1)}}=\rx + (1-\rx ) \,
\int_{\man} \, dY \kx \: \Qd{D}{k}{\Pe{Y}{\SD{n}{(0)}}}
\label{eq4.7}
\end{eqnarray}

\noindent Then, recalling that

\begin{equation} \label{eq4.7.1}
\Pe{X}{\SD{any \, level}{(0)}} = \rx
\end{equation}

\noindent we easily find

\begin{equation} \label{eq4.7.2}
\Pe{X}{\SD{0}{(n)}} =
\underbrace{F[F[\ldots F}_{n \, times}[\rx ]\ldots ]] = F_{n}[\rx ]
\end{equation}

\noindent Here $F_{n}[\phi(X)]$ denotes
the $n$-th iteration of the integral
operator

\begin{equation}
F[\phi(X)]=\rx + (1-\rx )\int_{\man} \, dY \kx \: \Qd{D}{k}{\phi(Y)}
\label{eq4.8}
\end{equation}

{}From Eqs.\ (\ref{eq4.2})-(\ref{eq4.8}) one can derive
that in the $n \rightarrow \infty $ ($t \rightarrow \infty $) limit
the
iterations converge (under some restrictions on \rx \, \, and \kx \,
which are necessary to guarantee the uniqueness of the fixed point)
to
the solution of the following integral equation

\begin{equation}
\Pe{}{X}=F[\Pe{}{X}]     \label{eq4.9}
\end{equation}

It is worth noting that after Eqs.\ (\ref{eq4.2})-(\ref{eq4.7}) it
becomes clear why we can substitute a $t$-dependent percolation
problem by $n$-dependent hierarchical one. Indeed, up to now we
didn't
mention that not all ``white'' cubes of given level break up
simultaneously and consequently there exist many ``white'' cubes
of different sizes at the time $t$. However, our approach is
insensitive to this difficulty because different cubes can join
the ``renormalization flow'' at different scales, but all of them
approach eventually the same fixed point (\ref{eq4.9}). The only
modification that can arise from this notion is the somewhat stronger
condition on the limiting rate $t \rightarrow \infty $ when one
proves the coincidence of $\lim_{t \rightarrow \infty }
\Pe{X}{\mbox{\cal B}_{t}} $
and \Pe{}{X} from (\ref{4.1.1}).

In some cases, when the full transition probability density \kx \,
(\ref{eq3.29})
can be approximated by a finite sum like

\begin{equation}
-n(Y) \sum_{s=1}^{J} \frac{1}{\kappa_{s}} K_{s}(X) \, K_{s}(Y)
\end{equation}

\noindent the corresponding nonlinear
integral equation (\ref{eq4.9}) reduces
to an algebraic one. This fact can be useful in applications
\cite{LinM91}\@.

Suppose, that we found the solution of (\ref{eq4.9}). Then the
solution
of the percolation problem is straightforward. The quantity

\begin{equation}
p_{\ast}=\int_{{\cal M}\backslash \Gamma} \gamma (X_{0}) \,
\Pe{}{X_{0}} \, dX_{0}          \label{eq4.10}
\end{equation}

\noindent has the meaning of the probability
that the cube of zero level is SD.
If $p^{(1)}_{D}$ is the percolation threshold for the site
percolation
problem on a cubic $D$-dimensional lattice $Z^{D}$, then, relating
black sites to SD cubes, white sites to non-SD cubes
and recalling the Proposition \ref{prop1}, we come immediately to

\begin{theorem}
The sufficient condition for percolation of \mbox{${\sf B}_{\infty}$}
is
\begin{equation}
p_{\ast} > p_{D}^{(1)}  \label{eq4.11}
\end{equation}                                        \label{th1}
\end{theorem}

The necessary condition for percolation is
the negation of the sufficient
condition for non-percolation. Let us introduce the notion of
``strictly closing'' (SC) configurations to obtain the sufficient
condition for non-percolation.

\begin{defi} \label{def2}
A cube of level $L$ is called ``strictly closing''
of zero rank \mbox{\rm (\SC{L}{(0)})} if it is white.
A cube is called ``strictly closing''
of rank $m$ \mbox{\rm (\SC{L}{(m)})}
if it is either white or
it consists of such configuration of \SC{L+1}{(m-1)} cubes that
there is no connected cluster of the remaining cubes of level $L+1$
belonging to \SC{L}{(m)} cube which could
connect any pair of faces of that cube. A cube is said to be \SC{}{}
if it is \SC{}{(n)} for some $n \geq 0$\@.
\end{defi}

One can easily verify the

\begin{prop}
Any number of \SC{}{} cubes, attached face to face, do not contain
any percolating path of black cubes. If the cube is \SC{}{(m)},
then it is \SC{}{(n)} for every $n>m$\@.
\label{prop2}
\end{prop}

Now it is clear that our ``renormalization'' procedure can be
applied.
Let us denote by the bar all quantities concerning the \SC{}{} cubes.
In complete analogy with the Eqs.\ (\ref{eq4.1})-(\ref{eq4.9}) we
obtain

\begin{eqnarray} \label{eq4.12}
\Pb{X}{\SC{0}{(n)}} = (1-\rx ) \,
\int_{\man} \, dY \kx \: \Qb{D}{k}{ \Pb{X}{\SC{1}{(n-1)}} }  \\
\vdots \nonumber   \\
\Pb{X}{\SC{n-1}{(1)}} = \underbrace{\Phi [\Phi [\ldots \Phi}_{n
\, times}[\Pb{X}{\SC{n}{(0)}}]\ldots ]] = \Phi_{n}[1-\rx ]
\label{eq4.13}
\end{eqnarray}

\noindent where we used $\Pb{X}{\SC{any \, level}{(0)}} = 1 - \rx $
and
the definitions

\begin{equation}
\Phi [\phi (X)] = (1-\rx ) \, \int_{\man} \, dY \kx \:
\Qb{D}{k}{\phi (Y)}  \label{eq4.14}
\end{equation}

\begin{equation} \label{4.14.1}
\Pb{}{X} = \lim_{n \rightarrow \infty} \Pb{X}{\SC{0}{(n)}}
\end{equation}

\noindent Then, we have (under the condition
that $\Phi[\phi(X)]$ has a unique
fixed point):

\begin{equation}
\Pb{}{X} = \Phi [\Pb{ }{X} ]  \label{eq4.15}
\end{equation}

\noindent Here \Qb{D}{k}{q} is an analog of \Qd{D}{k}{p} for closing
configurations ($q=1-p$ is the probability of white color,
p is the probability of black one). We will use in the next
Section only the simplest
combinatorial function of that type

\begin{equation}
\Qb{2}{3}{q} = q^{9} + 5q^{8}p + 10q^{7}p^{2} + 4q^{6}p^{3} +
q^{5}p^{4} = 3q^9 - 7q^8 + 4q^7 + q^5
\end{equation}

\Pb{}{X} is the probability for the
zero level cube to be SC\@. Associating white sites of
black site percolation problem on $Z^{D}$ to SC cubes,
black sites to non-SC cubes and
keeping in mind the Proposition \ref{prop2}, we come to

\begin{theorem}
Let $p_{D}^{(2)} = 1-q_{D}^{(2)}$ be the non-percolation
threshold for the black
site percolation problem on the $Z^{D}$\@. Denote
\label{th2}

\begin{equation}
q_{\ast } =\int_{{\cal M}\backslash \Gamma}{ \gamma (X_{0}) \:
 \Pb{}{X_{0}} \, dX_{0}}
\label{eq4.16}
\end{equation}

\noindent Then the necessary condition for
percolation (negation of the sufficient
condition for non-percolation) of
\mbox{${\sf B}_{\infty}$} is

\begin{equation}
q_{\ast } < q_{D}^{(2)}         \label{eq4.17}
\end{equation}
\end{theorem}

\section{THE SIMPLEST MODEL}            \label{sec5}

Here we will investigate the simplest model of the type described
above.
It corresponds to homogeneous diffusion on the line segment
${\cal M}=[0;l] ; \Gamma = \{ 0;1\} ; d=1 $  with

\begin{equation}
{\cal F}(x) = 0 \; ; \; \sigma (x) = \sigma =const  \label{eq5.1}
\end{equation}

\begin{equation}
n(x) = a^{2} = const \; ; \; \gamma (x) = \frac{1}{l}
\label{eq5.2}
\end{equation}

\begin{equation}
\hat{A} \, f(x) = \frac{\sigma^{2}}{2} \frac{d^{2}}{dx^{2}} f(x)
\label{eq5.3}
\end{equation}

\noindent and to $D=2; k=3; r=9 $ case of the cubic net.
{}From Eqs.\ (\ref{eq3.12}),(\ref{eq3.13}) we obtain

\begin{equation}
\lambda_{1} = 8a^{2} - \frac{\pi^{2} \sigma^{2} }{2l^{2}}
\label{eq5.4}
\end{equation}

\noindent and the non-degeneracy condition is
$\frac{al}{\sigma } > \frac{\pi }{4} $\@.
Assuming this, one can derive from equations
(\ref{eq3.19}), (\ref{eq3.24})-(\ref{eq3.29})

\begin{equation}
\rho(z) = \frac{\cosh(\frac{1}{2}\zeta - z)}{\cosh(\frac{1}{2}\zeta)}
\label{eq5.5}
\end{equation}

\begin{equation}
K(z,u) = \left\{ \begin{array}{lc}
   \mbox{$\frac{\zeta}{l \sinh(\zeta)}
\sinh(\zeta -z) \, \sinh(u) , $} &
   \mbox{if $z \geq u$}  \\
   \mbox{$\frac{\zeta}{l \sinh(\zeta)} \sinh(\zeta -u) \,
\sinh(z) , $}  &
   \mbox{if $z \leq u$}
   \end{array}   \right.      \label{eq5.6}
\end{equation}

\noindent where we have introduced the dimensionless variables
and the parameter $\zeta$, which is in fact the only dimensionless
parameter in this model and on which the behaviour of the model
depends essentially

\begin{equation}
z = \frac{\sqrt{2} ax}{\sigma} \; \; ; \; \;
u = \frac{\sqrt{2} ay}{\sigma} \; \; ; \; \;
\zeta = \frac{\sqrt{2} al}{\sigma} \label{eq5.6.1}
\end{equation}

\noindent Equation (\ref{eq4.9}) now reads

\begin{eqnarray}
 & & \Pe{}{z} = \frac{\cosh(\frac{1}{2}\zeta -
z)}{\cosh(\frac{1}{2}\zeta)}  +
\left( 1-\frac{\cosh(\frac{1}{2}\zeta - z)}{\cosh(\frac{1}{2}\zeta)}
\right) \left[ \sinh(\zeta -z) \,
\int_0^z \, du \, \frac{\sinh(u)}{\sinh(\zeta)} \,
\Qd{2}{3}{\Pe{}{u}} \right. \nonumber \\
 & & \mbox{ }+ \left. \sinh(z) \,
\int_z^{\zeta} \, du \, \frac{\sinh(\zeta -u)}{\sinh(\zeta)} \,
\Qd{2}{3}{\Pe{}{u}} \right]
     \label{eq5.7}
\end{eqnarray}

\noindent Introduce the function

\begin{equation}
\beta(z) =
\int_0^z \, du \, \frac{\sinh(u)}{\sinh(\zeta)} \,
\Qd{2}{3}{\Pe{}{u}}     \label{eq5.8}
\end{equation}

\noindent Because we have $\rho(\zeta -z) = \rho(z) $,
symmetry reasons lead us
to $\Pe{}{\zeta -z} = \Pe{}{z} $ and Eq.\ (\ref{eq5.7}) now reads

\begin{equation}
\Pe{}{z} = \frac{\cosh(\frac{1}{2}\zeta -
z)}{\cosh(\frac{1}{2}\zeta)}  +
\left( 1-\frac{\cosh(\frac{1}{2}\zeta -
z)}{\cosh(\frac{1}{2}\zeta)}\right)
\left( \sinh(\zeta -z) \, \beta(z) + \sinh(z)
\, \beta(\zeta -z) \right)
 \label{eq5.9}
\end{equation}

One can derive an ordinary differential equation for $\beta(z)$
from (\ref{eq5.8}), (\ref{eq5.9}), but it is too involved and we
don't produce it here. Instead, we present the result of the computer
solution of inequality (\ref{eq4.11}) in terms of $\zeta$, which
was calculated using (\ref{eq5.8}), (\ref{eq5.9}) and assuming
the known value of $p_{2}^{(1)} = 0.59 $\@. Thus, the sufficient
condition is

\begin{equation}
\zeta \leq \zeta_{suff} = 3.29      \label{eq5.10}
\end{equation}

\noindent Equations for the necessary condition are
(see (\ref{eq4.14})-(\ref{eq4.17}))

\begin{equation}
\Pb{}{z} = \left( 1-\frac{\cosh(\frac{1}{2}\zeta - z)}
{\cosh(\frac{1}{2}\zeta)}\right)
\left( \sinh(\zeta -z) \, g(z) + \sinh(z) \, g(\zeta -z) \right)
\label{eq5.11}
\end{equation}

\begin{equation}
g(z) =  \int_0^z \, du \frac{\sinh(u)}{\sinh(\zeta)} \,
\Qb{2}{3}{\Pb{}{u}}
\label{eq5.12}
\end{equation}

Our estimate for the necessary condition
(\ref{eq4.17}) is too crude for
this particular model and we don't produce the result here.
However, we
think that the sufficient condition~(\ref{eq5.10})
is reasonably close to the exact
critical value of $\zeta $\@. This can be confirmed
by calculation of the next level corrections to it (compare with
Eq. (\ref{a1}), (\ref{a6}), where
such corrections are the leading ones).

\section{SUMMARY AND DISCUSSION}                \label{sec6}

We have described a new class of problems, synthesizing
ones from familiar percolation and branching diffusion. To find
the percolation characteristics, one has to solve the analytic
relations (\ref{eq4.8})-(\ref{eq4.11}) and
(\ref{eq4.14})-(\ref{eq4.17})\@.

Let us emphasize some possible generalizations. First of
all, it is worth noting that face-connectedness, which we
assumed throughout this paper, can be replaced by a great variety
of other ``quasilocal'' definitions of connectedness. By this
we mean a definition, which would deal with some finite clusters
of black cubes (possibly of different sizes) with some rules
restricting their configurations. If something like
Propositions~\ref{prop1},~\ref{prop2} holds, then
only minor modification of our approach, reducing mainly to
change of combinatorial functions \Qd{D}{k}{p} and \Qb{D}{k}{q}
and, maybe, the number of levels of hierarchy involved in
construction of recurrent formulas of type (\ref{eq4.2}),
(\ref{eq4.12}) should be expected. For example, one can
consider two black cubes to be connected iff the minimum path
from one to the other contains no more than $R_{0}$ white cubes.
This would correspond, in some sense, to defects with ``spheres of
influence'', which were investigated in some one-scale
models~\cite{MMS86} ($R_0 = 0$ corresponds to the face to face
connectedness, i.e.\  the present paper)\@.
With this modification, the problem will have an additional
parameter $R_0$ and, presumably,  non-trivial dependence
on that parameter. One can consider a new problem taking the
other parameters of the model fixed and letting $R_0$ vary.
It is clear that  even if there were no percolation
of this ``spheres of influence'' for small $R_0$, it would presumably
occur for safficiently large ``radius of sphere'' $R_0$, and,
therefore,
we would have a percolation phase transition at some intermediate
value of $R_{0}$\@.

However, the consideration of more than two levels of
hierarchy appears to be necessary already for $R_0 = 0$
if we consider $k=2$ case.
Here we present the analogs of the equations (\ref{eq4.8})
and (\ref{eq4.14}) for the $k=2$ case. Because at the first step of
breaking up there are no nontrivial SD (and SC) configurations,
one has to consider two steps of breaking up (and, therefore,
three levels of hierarchy) in order to derive the corresponding
integral operator. The form of this integral operator is given by

\newcommand{\dry}[1]{\mbox{$\rho^{#1}(Y)$}}
\newcommand{\dyr}[1]{\mbox{$(1-\rho (Y))^{#1}$}}
\newcommand{\dQa}[1]{\mbox{$Q^{(1)}_{2,2}(#1)$}}
\newcommand{\dQb}[2]{\mbox{$Q^{(2)}_{2,2}(#1,#2)$}}
\newcommand{\dQc}[3]{\mbox{$Q^{(3)}_{2,2}(#1,#2,#3)$}}
\newcommand{\dQd}[4]{\mbox{$Q^{(4)}_{2,2}(#1,#2,#3,#4)$}}
\newcommand{\dQn}[4]{\mbox{$
\overline{Q}^{(4)}_{2,2}(#1,#2,#3,#4)$}}

\begin{eqnarray}
 & & F^{(2)}[\Pe{}{X}] = \rx + (1-\rx ) \int dY \kx [ \dry{4}
\nonumber  \\
 & & \mbox{ }+ \dry{3} \dyr{} \int dZ K(Y,Z) \dQa{\Pe{}{Z}}
\nonumber    \\
 & & \mbox{ }+ \dry{2} \dyr{2} \int dZ dU K(Y,Z) K(Y,U)
\dQb{\Pe{}{Z}}{\Pe{}{U}} \nonumber   \\
 & & \mbox{ }+ \dry{} \dyr{3} \int dZ dU dV K(Y,Z) K(Y,U) K(Y,V)
\dQc{\Pe{}{Z}}{\Pe{}{U}}{\Pe{}{V}} \nonumber   \\
 & & \mbox{ }+ \dyr{4} \int dZ dU dV dW K(Y,Z) K(Y,U) K(Y,V) K(Y,W)
\nonumber \\
 & &\times \dQd{\Pe{}{Z}}{\Pe{}{U}}{\Pe{}{V}}{\Pe{}{W}} ]
\label{a1}
\end{eqnarray}

\noindent where the combinatorial functions \dQa{p_1},
\dQb{p_1}{p_2},
\dQc{p_1}{p_2}{p_3}, \dQd{p_1}{p_2}{p_3}{p_4} are introduced
which count the ``strongly defective'' configurations of black
cubes of two subsequent levels $L+1$ and $L+2$ of hierarchy
within the ``parent'' cube of level $L$ for the cases when only
one (two, three or four, respectively) of the cubes of level $L+1$
is broken up to cubes of level $L+2$ (and the first term
in the integrand takes into account the only SD
configuration when none of $L+1$ level cubes
is broken up)\@. Their expressions as polynomials of $p_{i}$
are too involved and do not seem to contain any clarifying
information, so we do not produce them here.
Equations (\ref{eq4.9})-(\ref{eq4.11}) are implemented
with  $F^{(2)}[\cdot ]$  instead of $F[\cdot ]$ \@.
Analogously, for the necessary condition we should use in
(\ref{eq4.15})  the following expression instead of
$\Phi [\cdot ]$ (see (\ref{eq4.14}))

\begin{eqnarray}
 & & \Phi^{(2)}[\Pe{}{X} ] = (1-\rx )\, \int dY \kx [ \dyr{4}
\nonumber \\
 & & \mbox{ } \times \int dZ dU dV dW K(Y,Z) K(Y,U) K(Y,V) K(Y,W)
\nonumber \\
 & & \mbox{ } \times \dQn{\Pe{}{Z}}{\Pe{}{U}}{\Pe{}{V}}{\Pe{}{W}} ]
\label{a6}
\end{eqnarray}

\noindent with obvious notations.

Second, one can generalize the branching diffusion part of
the problem. For instance, the number $r$ \, can be considered
to be random (possibly as a function of $X_{t}$ )\@. This, of
course, will make the problem of definition of ``strongly defective''
and ``strictly closing'' configurations more complicated because
in this case we should deal with face to face attached cubes which
are broken up in substantially different ways and we should pay
much more attention to ensuring the existence of something like
Propositions \ref{prop1}, \ref{prop2}\@. However, we think that
the revision might be only technical (taking into account that
the branching diffusion part of the problem may be generalized
to this case easily \cite{Berest} )
and in physical applications such a modification can always be
absorbed by some corrections to the function $n(X)$\@.

One can improve the evaluation of both necessary and sufficient
conditions considering three (or more) steps instead of two.
In principle this can be done and one would obtain the equations
similar to (\ref{a1}) and (\ref{a6}), but computational difficulties
increase very fast making this method hardly applicable.

For applications it is very interesting to develop any
method of approximate solution of equations
(\ref{eq4.9}), (\ref{eq4.15})\@. Sometimes, approximations, which
reduce \kx \, \, to degenerate kernel, are not so bad.

Anyway, the investigated problem has physical significance
\cite{LinM91}\@. It may occur that slightly complicated
models of fracture, currently used in geophysics
\cite{MPR87} -\cite{ALMP87}
as well as some generalizations of the models currently used
to describe the intermittent behaviour of high-energy scattering
processes \cite{IntermHEC}
also will lead to similar problems. We hope that the applications
will not be limited to these known areas.

\vspace{0.5cm}

A.M. is indebted to A.D.Linde for suggesting the physical
problem \cite{LinM91}, that led us finally to the present
work. He is also grateful to A.A.Starobinsky, Ya.G.Sinai,
L.Bogachev and O.Filippova for useful discussions on various
aspects of this work. We highly appreciate the careful and
kind reference and are grateful to the referee for very
useful suggestions concerning especially the definitions of
\SD{}{} and \SC{}{} cubes.


\listoffigures

\newpage

\newpage

\end{document}